\documentclass{article}
\usepackage[T1]{fontenc} % add special characters (e.g., umlaute)
\usepackage[utf8]{inputenc} % set utf-8 as default input encoding
\usepackage{ismir,amsmath,cite,url}
\usepackage{graphicx}
\usepackage{booktabs}

\usepackage{color}

\usepackage[firstpage]{draftwatermark}
\definecolor{lightgray}{rgb}{0.9,0.9,0.9}
\definecolor{darkgray}{rgb}{0.4,0.4,0.4}
\SetWatermarkFontSize{12pt}
\SetWatermarkScale{1.1}
\SetWatermarkAngle{90}
\SetWatermarkHorCenter{202mm}
\SetWatermarkVerCenter{170mm}
\SetWatermarkColor{darkgray}
\SetWatermarkText{Late-Breaking / Demo Session Extended Abstract, ISMIR 2024 Conference}

\def\lbd{}	% Flag to use correct LBD settings in the paper, please do not modify this line

\title{Hookpad Aria: A Copilot for Songwriters}

\multauthor
{Chris Donahue$^{1}$ \hspace{1cm} Shih-Lun Wu$^{2}$ \hspace{1cm} Yewon Kim$^3$} { \bfseries{Dave Carlton$^{4}$ \hspace{1cm} Ryan Miyakawa$^{4}$ \hspace{1cm} John Thickstun$^{5}$}\\
 $^{1}$ CMU, $^{2}$ MIT, $^{3}$ KAIST, $^{4}$ Hooktheory, $^{5}$ Cornell
}

\def\authorname{C. Donahue, S.-L. Wu, Y. Kim, D. Carlton, R. Miyakawa, and J. Thickstun}

\usepackage[bookmarks=false,pdfauthor={\authorname},pdfsubject={\papersubject},hidelinks]{hyperref}
\usepackage{cleveref}

\newcommand\sysname{Hookpad Aria}
\newcommand\hookpad{Hookpad}

\begin{document}

\maketitle

\url{https://hooktheory.com/hookpad/aria}

\begin{abstract}
We present \sysname, a generative AI system designed to assist musicians in writing Western pop songs. 
Our system is seamlessly integrated into \hookpad{}, a web-based editor designed for composition of \emph{lead sheets}---symbolic music scores that describes melody and harmony. 
\sysname{} has numerous generation capabilities designed to assist users in non-sequential composition workflows, including: 
(1)~generating left-to-right continuations of existing material, 
(2)~filling in missing spans in the middle of existing material, 
and 
(3)~generating harmony from melody and vice versa. 
%Across all capabilities, our system can incorporate global context about a user's composition: meter, key, and tempo. 
%The goal of our system is to take the first step towards a future where musicians directly benefit from generative AI models trained at scale, in a manner analogous to that of programmers benefiting from code generation. 
\sysname{} is also a scalable data flywheel for music co-creation---since its release in March 2024, Aria has generated $318$k suggestions for $3$k users who have accepted $74$k into their songs.
\end{abstract}

\vspace{-3mm}
\section{Introduction}\label{sec:introduction}

% Advancements in generative AI for music

%\todo{writing down misc thoughts that could go here. need to be organized into a story}
% \yewon{One suggestion is to combine abstract + intro (unless abstract is explicitly required) and start directly by introducing Aria and highlighting its key functionalities. I think beginning with a broader storyline—starting from AI advancements, then covering AI creative support systems, followed by the shortcomings in the music domain and limitations in music AI systems, and finally differentiating Aria—would be too lengthy for an LBD paper.} % nevermind, i guess ISMIR expects the predefined paper structure

% % 
% % Recent advancements in AI transforms human creative process(?)
% % accelerate integration of AI in creative process in various domains
% While AI playing crucial role in creative tasks in various domains~\cite{}, music is relatively slow

% % However, music is relatively undeexplored
% Despite recent dramatic improvements in the capabilities of gen AI for  music in both ...., 

% Recent and dramatic improvements in the capabilities of generative AI for music in both its  symbolic
\noindent
%\slwu{
Recently, a growing body of research on generative models has 
%equipped machines with 
yielded 
impressive music generation capabilities in both
%compose and/or perform music in both 
%generate music in both 
symbolic~\cite{simon2017performance,huang2018music,donahue2019lakhnes,payne2019musenet,thickstun2023anticipatory,wu2023compose} and acoustic~\cite{oord2016wavenet,dieleman2018challenge,dhariwal2020jukebox,forsgren2022riffusion,agostinelli2023musiclm,donahue2023singsong,copet2024simple,wu2023music} domains. 
These models offer tremendous potential to enrich and accelerate music creation via integration with existing workflows.
%}

% ~\cite{simon2017performance,huang2018music,donahue2019lakhnes,payne2019musenet,thickstun2023anticipatory} and acousticforms.
% Massive potential for this work to transform music creation workflows.

%\slwu{
Beyond the music domain, generative models have been integrated into various creative workflows, including video editing~\cite{wang2024lave}, 3D modeling~\cite{liu20233dalle}, story writing~\cite{wordcraft}, and programming~\cite{mcnutt2023design}. 
% Integration of generative models into creative workflows is increasingly prevalent across domains like 
% video editing~\cite{wang2024lave}, 
% %visual design~\cite{promptpaint}, 
% 3D modeling~\cite{liu20233dalle}, 
% creative writing~\cite{wordcraft}, 
% and programming~\cite{mcnutt2023design}. 
%Programming in particular is perhaps the most mature of these integrations---
%and 
%text (e.g., story writing~\cite{wordcraft} and programming~\cite{mcnutt2023design}).
%User studies on GitHub Copilot~\cite{copilot}, 
% Such integrations int
%Integrations with programming workflows are perhaps the most mature
These integrations have been shown to support the creation process effectively---studies with users of GitHub Copilot, a prominent AI programming assistant, indicate that the tool not only enhances productivity but also enjoyment and satisfaction in the process~\cite{kalliamvakou2022copilot}.
%suggested that it increases not only programmers' productivity, but also their satisfaction~\cite{kalliamvakou2022copilot}.
% CHRIS: I don't think we need to speculate as to why this hasn't happened yet, just say we're starting to do it now
%However, this integration process has been relatively slow for music, potentially due to the difficulty of deploying generative models in digital audio workstations (DAWs), an environment which musicians are familiar with~\cite{beckerdesigning}.
%}

Hoping to offer similar benefits to musicians, 
who are currently (and understandably) skeptical about AI~\cite{lovato2024foregrounding}, 
here we aim to integrate generative models into the workflows of songwriters. 
Specifically, we build a Copilot-like system called \sysname{}, 
an integration of state-of-the-art symbolic music generation capabilities into \hookpad{}, 
a popular web editor designed for songwriting (in the last month, $7$k \hookpad{} users wrote $33$k songs). 
\hookpad{} supports the composition of lead sheets, 
symbolic musical scores that describe melody and harmony---the essence of a piece of Western pop music. 
Our system aims to seamlessly integrate with \hookpad{} to support the specific needs of songwriters on that platform. 
In contrast, 
past work proposes new workflows~\cite{huang2019bach,louie2020cococo} rather than integrating with existing ones, 
or focuses on different platforms~\cite{hadjeres2021piano,malandro2023composer,becker2024designing}. 
Unlike past work, 
we also aim to collect interaction data from users, 
with the goal of supporting future research efforts. 
Users explicitly opt in to data collection when they first use \sysname---to date we have collected implicit feedback on $318$k suggestions from $3$k unique users.
Additionally, we report findings from interviews with eight \sysname{} users to gain insights into their experiences, perceptions, and how they engage with the tool.

\vspace{-3mm}
\section{Hookpad Aria}

\begin{figure*}[t]
    \centering
    \includegraphics[width=0.99\linewidth]{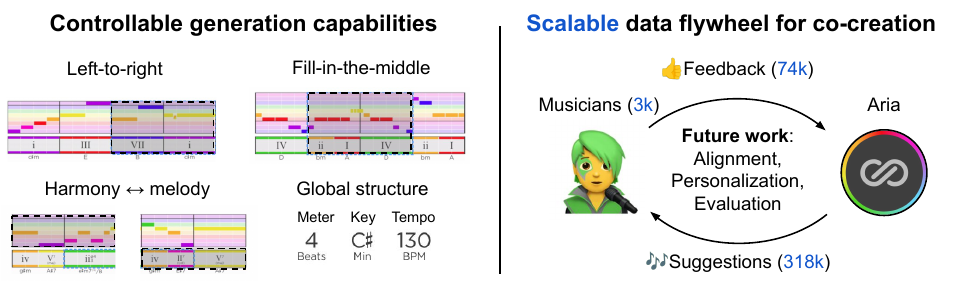} % Change the width as needed
    \caption{\sysname{} supports human-AI co-creation of pop songs through flexible control capabilities (shaded regions are output): left-to-right, fill-in-the-middle, harmony to melody, melody to harmony, all conditioned on global structure attributes (meter, key, tempo). \sysname{} represents a scalable data flywheel for music co-creation. Since March 2024, Aria has generated $318$k suggestions for $3$k unique users---users have explicitly accepted $74$k suggestions into their songs.}
    \label{fig:aria}
\end{figure*}

%Here we describe our system design and LLM backbone.

%\vspace{-3mm}
\textbf{System design.}
% (1)~supporting non-sequential workflows,
% Non-sequential workflows supported through unprecedented control capabilities for symbolic generation.
% In addition to standard left-to-right generation, we do fill-in-the-middle, harmonization, melodic accompaniment.
% All sensitive to a user's local context, as well as global structure such as meter, key, and tempo.
% Past work supports some of this but not all.
The design of \sysname{} centers around two key principles. 
Firstly, we aim to support songwriters in \emph{non-sequential} workflows. 
Accordingly, we designed our system to offer unprecedented control capabilities for symbolic music generation. 
In addition to 
sequential left-to-right generation, 
we support fill-in-the-middle, 
harmonization from melody, and
generating melodies from harmony (\Cref{fig:aria}).  
Past work on symbolic music generation offers some of these features but not all~\cite{huang2018music,huang2019bach,louie2020cococo,hadjeres2021piano,yeh2021automatic}. 

%(2)~seamless integration with Hookpad. 
% Seamless integration.
% Users highlight spans of time in the Hookpad interface, repurpose familiar UI for selecting regions of the song.
% Users can select to generate melody, harmony, or both for that time span, conditioned on all surrounding context. 
% User receives several suggestions and can audit them one at a time. 
% Users then ``accept'' their favorite which is committed to their tracks.
% We collect implicit feedback data in the form of which suggestions users accepted or ignored.
Secondly, 
inspired by the seamless integrations of tools like Github Copilot into code editors, 
we aim to seamlessly integrate our system into \hookpad.
We repurpose the familiar measure selection workflow to allow users to select regions of a song to generate. 
Users specify if they want to generate melody, harmony, or both for the selected span, conditioned both on surrounding local context as well as global attributes (meter, key, and tempo) of their project. 
\sysname{} displays endless alternative generations from Aria for a single span, which users can audit until finding one they wish to ``accept'' into their track. 
We log whether suggestions are accepted or ignored as implicit feedback data---$74$k have been accepted to date.

\vspace{2mm}
\noindent \textbf{LLM backbone.}
%\subsection{LLM backbone}
%\chris{this is the only section besides abstract where I've written any prose and it's very much a work in progress}
\sysname{} is based on the Anticipatory Music Transformer~\cite{thickstun2023anticipatory}, a state-of-the-art large language model pre-trained for multi-instrumental symbolic music generation.
The Anticipatory Music Transformer is capable of both standard autoregressive (left-to-right) generation conditioned on preceding context, 
and also has the capability to fill-in-the-middle based on both preceding and subsequent context. 
% \emph{infill}, 
% i.e.,~generate conditioned on both preceding and subsequent context. 
To adapt this model to our setting, we fine tune the ``medium'' pre-trained model ($360$m parameters) on $50$k lead sheets from TheoryTab.\footnote{\url{https://www.hooktheory.com/theorytab}}

In the Anticipatory Music Transformer,
MIDI notes (consisting of absolute start time, duration, instrument category, and pitch) 
are partitioned into two sequences:
events~$\mathbf{e}$ and controls~$\mathbf{c}$. 
Controls are shifted forward in time by $5$ seconds and interleaved with events into a single sequence which is then modeled with a standard Transformer LM, representing ${P_{\theta}(\mathbf{e} \mid \mathbf{c})}$. 
This setup allows the model to ``anticipate'' control notes up to $5$ seconds in advance of their arrival. 
By partitioning training examples such that notes in subsequent context are assigned to controls, 
the Anticipatory Music Transformer learns to fill-in-the-middle.

% This setup is meant to facilitate unusually versatile control for music generation, allowing for generating notes from any other sequence of notes (e.g., generating melody from harmony, or generating the past from the future). 
% Accordingly, 
% the events and control sequences can have arbitrary timings (e.g.,~they can overlap in time, or the controls can come after the events). 
% To make this tractable to model, the Anticipatory Music Transformer adopts a factorization that allows the model to ``anticipate'' controls some number of seconds $\delta$ into the future:
% \begin{align*}
% P_\theta(\mathbf{e}|\mathbf{c}) &= \prod_{i=1}^{N} P_\theta(e_i|\mathbf{e}_{<i}, \mathbf{c}_{\mathcal{I}_{i}}),~\text{where} \\
% \mathcal{I}_{i} &= \{j \mid 1 < j \leq M, c_j\notetime < e_i\notetime + \delta\}.
% \end{align*}
% In both this work and the original paper, $\delta = 5$ seconds.

There are two challenges in adapting the Anticipatory Music Transformer to our setting. 
Firstly, \hookpad{} uses a proprietary representation of lead sheets rooted in functional harmony. 
Here we design an encoding scheme to convert between the functional representation and corresponding MIDI notes. 
Secondly, \hookpad{} indexes time in units of beats, rather than absolute time. 
To represent beats in absolute time, 
we add a click track ``instrument'' with one note per beat, 
%we convert tempo into a click track ``instrument'' with one note per beat, 
allowing the model to anticipate clicks and learn to output notes in lockstep.  

A lead sheet thus consists of three sets of notes comprising the melody $\mathcal{M}$, harmony $\mathcal{H}$, and click track $\mathcal{C}$. 
For each song in the Theorytab data, 
we construct random fine tuning examples by:
(1)~selecting a random time span $[t_s, t_e]$ corresponding to a user selection,
(2)~randomly choosing a control capability we aim to support, and
(3)~partitioning the song into events and controls based on the selected capability--see \Cref{tab:partitions}. 
Our backbone is the result of fine tuning on many of these random examples per song.
%We create many of these random examples per song and use them to fine tune, resulting in the backbone deployed in the live system.

\begin{table}[t]
\centering
\begin{tabular}{ccc}
\toprule
Capability          & Events ($\mathbf{e}$)                                       & Controls ($\mathbf{c}$)                                                  \\ \midrule
Left-to-right & $\mathcal{M} \cup \mathcal{H}$ & $\mathcal{C}$ \\
Fill-in-middle & $\mathcal{M}_{< t_e} \cup \mathcal{H}_{< t_e}$ & $\mathcal{M}_{\geq t_e} \cup \mathcal{H}_{\geq t_e} \cup \mathcal{C}$ \\
Harm-to-mel & $\mathcal{M}_{< t_e} \cup \mathcal{H}_{< t_s}$ & $\mathcal{M}_{\geq t_e} \cup \mathcal{H}_{\geq t_s} \cup \mathcal{C}$ \\
Mel-to-harm & $\mathcal{M}_{< t_s} \cup \mathcal{H}_{< t_e}$ & $\mathcal{M}_{\geq t_s} \cup \mathcal{H}_{\geq t_e} \cup \mathcal{C}$ \\
\bottomrule
\end{tabular}
\caption{We create fine tuning data to support different infilling capabilities by partitioning notes of melody ($\mathcal{M}$), harmony ($\mathcal{H}$), and click track ($\mathcal{C}$) into events and controls for anticipation based on random time spans $[t_s, t_e]$.}
\label{tab:partitions}
\vspace{-4mm}
\end{table}

\vspace{-3mm}
\section{Findings from Aria users}

% Stats from Ryan

% how many outputs have been generated with Aria
% 318408

% how many scenarios have been encountered by Aria (unique contexts for which multiple outputs may hav been generated)
% only thing not immediately queryable, will look into this

% how many unique users have tried Aria
% 3242

% how many “power users” do we have (>100 scenarios?)
% 449

% what % of Aria outputs have been accepted
% 23.3%

%Power users: $449$ have generated $>100$ model outputs.

%We summarize the major themes of finding from an interview study.
% in the intro, say Through a 1-hour interview study with eight users, we studied how users use Aria and their perceptions and experiences.
Here we summarize findings from $1$-hour semi-structured interviews with eight \sysname{} users. 

\textbf{Aria facilitates ideation.} Participants viewed \sysname{} as a fellow songwriter, turning to it when they encountered creative blocks. For example, they used it to get a starting point to riff off or connect two melodic phrases. While they rarely used the suggestions verbatim, the co-ideation process often helped them maintain their creative momentum, preventing fatigue from creative block.

% viewed Aria as a 
% collaborator, especially useful for overcoming writer's block. 
% For example, they used Aria to generate melodies when they lacked a clear starting point. 
% %, often drawing inspiration from the rhythm or shape of the generated melody. Participants also turned to Aria to fill in specific parts of their songs, such as creating a pre-chorus given verse and chorus. %, or concluding melodic phrases.
% Although they seldom used Aria's generations without modification, they often found its suggestions helpful for ideation nonetheless.
% %and refining ideas that aligned with their creative vision.
% % mention their original process (improvisation with instruments) and how using Aria saves time and effort - probably analogous to code writing assistants? mention them if there are similar findings

\textbf{Aria preserves agency.} 
Users remarked that three aspects of \sysname{} conveyed a sense of agency. 
Firstly, its short, reusable suggestions, described as creative sparks, allowed for greater autonomy compared to text-to-music generation tools, which often diminished users' agency and joy of creation.
Second, its seamless integration into the Hookpad editor allowed users to modify the suggestions as they saw fit. 
Finally, they appreciated its non-sequential generation, offering them full control over how, when, and where to use the system.

\textbf{More control is needed.}
While Aria offers both local and global forms of control, 
participants expressed a desire for even \emph{more} control. 
%, beyond the existing control based on local context and global structure. 
%While Aria generates suggestions based on both the user's local context and the global structure of the song, participants expressed a desire for more control over its outputs. 
Specific forms of additional control mentioned included 
genre (e.g.,~ambient, rock), 
emotional tone (e.g.,~happy, sad), 
intended instruments (e.g.,~vocal, guitar), 
and structural elements (e.g.,~verse, chorus).
%Additionally, some participants mentioned the need for instrument-specific melodies, allowing them to tailor compositions for a guitar, violin, or vocalist. 
%For vocal melodies, they stressed the importance of singability, avoiding impractical features like rapid octave shifts.
% may link to "aligning models to user feedback / personalizing models" in research opportunities

\vspace{-3mm}
\section{Research opportunities}

The scalable data flywheel created by Aria offers many potential avenues for future research, 
such as aligning models to user feedback~\cite{ouyang2022training}, 
A/B testing different generative models for in-the-wild evaluation~\cite{chiang2024chatbot}, 
and producing interaction datasets that help us to better study the nature of human-AI music co-creation~\cite{lee2022coauthor}. 
We plan to explore these in future work.

% Aligning models to user feedback (RLHF).
% Personalizing models (RLHF).
% In-the-wild eval (Chatbot Arena, A/B testing). 
% Understanding human-AI music co-creation (data science to understand scenarios where users reach for AI models and what they're hoping to get, releasing datasets like Mina's Co-author w/ users who opt in for community study)

% For bibtex users:
\bibliography{ISMIRtemplate}

\end{document}